# Ultrafast and Large Third-order Nonlinear Optical Properties of CdS Nanocrystals in Polymeric Film

**J. He, W. Ji,\* G. H. Ma, and S. H. Tang**
*Department of Physics, National University of Singapore, 2 Science Drive 3, Singapore 117542, Republic of Singapore*

**E. S. W. Kong**
*Nanophotonic Semiconductors, Singapore Science Park, Singapore 118633, Republic of Singapore*

**S. Y. Chow and X. H. Zhang**
*Institute of Materials Research and Engineering, Singapore 117602, Republic of Singapore*

**Z. L. Hua and J. L. Shi**
*Shanghai Institute of Ceramics, Chinese Academy of Sciences, Shanghai 200050, People's Republic of China*

**Abstract**

We report the ultrafast and large third-order nonlinear optical properties of CdS nanocrystals (NCs) embedded in a polymeric film. The CdS NCs of 2-nm radius are synthesized by an ion exchange method and highly concentrated in the two layers near the surfaces of the polymeric film. The two-photon absorption coefficient and the optical Kerr coefficient are measured with laser pulses of 250-fs duration at 800-nm wavelength. The one-photon and two-photon figures of merit are determined to be 3.1 and 1.3, respectively, at irradiance of 2 GW/cm$^2$. The observed nonlinearities have a recovery time of ~ 1 ps. The two-photon-generated free carrier effects have also been observed and discussed. These results demonstrate that CdS NCs embedded in polymeric film are a promising candidate for optical switching applications.

\* To whom correspondence should be addressed. Fax: (65) 67776126. Phone: (65) 68746373. E-mail: phyjiwei@nus.edu.sg.

## 1. Introduction

Semiconductor nanocrystals (NCs) have attracted increasing attention because of their tunable optical properties arising from the quantum size effect.[1–3] In particular, research efforts have been focused on the third-order, nonlinear-optical properties of semiconductor NCs, which may lead to potential applications in ultrafast all-optical switching.[1,3–7] For realization of all-optical switching devices based on waveguide structures with exploiting nonlinear phase changes, the following material requirements have to be met: $W > 1$ and $T < 1$, where $W$ and $T$ are the one- and two-photon figures of merit, respectively.[8] The two figures of merit are defined as $W = n_2 I/(\alpha\lambda)$ and $T = \beta\lambda/n_2$, where $n_2$ is the optical Kerr coefficient, $\alpha$ the absorption coefficient, $\beta$ the two-photon absorption coefficient, $\lambda$ the wavelength, and $I$ the light irradiance. Furthermore, ultrafast response times (a few picoseconds or less) are required for the nonlinear processes involved. Although resonant, third-order nonlinearities are large, they are generally slow and associated with large absorption. Hence, research attention has been paid to the nonlinear responses of II-VI semiconductor NCs in the spectral region having photon energies below the fundamental absorption edge.[1,3–7] In this case, off-resonant nonlinearities are dominated by two-photon absorption, optical Kerr nonlinearity, or two-photon-generated free carrier effects.[1,3–7] However, Banfi et al.[3] have found that the figures of merit are far away from the target values for glasses doped with semiconductor $CdS_{1-x}Se_x$ NCs.

To improve the figures of merit, research interest has been directed towards semiconductor NCs embedded in polymer. Schwerzel et al.[9] and Du et al.[10] have measured the nonlinear index of refraction in CdS NCs embedded in polymers with nanosecond laser pulses at wavelengths close to the fundamental absorption edge.

Recently, Lin *et al.*[11] have reported a large optical Kerr coefficient of $-8.4 \times 10^{-14}$ $cm^2/W$ for CdS NCs in poly(methyl methacrylate) (PMMA) with femtosecond laser pulses at 815 nm. They have also determined $W$ to be 1.2 for 3 wt% CdS/PMMA hybrid film.

Here we report our investigation into the ultrafast and large off-resonant, third-order, nonlinear responses of CdS NCs embedded in a polymeric film. The CdS polymeric film has been synthesized by an ion-exchange process, and characterized with various techniques like transmission electron microscopy (TEM), X-ray diffraction, transmission spectroscopy, photoluminescence (PL), and ellipsometry. With 250-fs laser pulses at 800-nm wavelength, we have performed a femtosecond time-resolved excite-probe measurement on the CdS polymeric film. In addition, we have carried out an experiment on femtosecond time-resolved optical Kerr effect. Our results show that the CdS-polymer composite film possesses a recovery time of ~ 1 ps with $W$ = 3.1 and $T$ = 1.3. The two-photon-generated free carrier effects have also been observed and discussed. Our findings indicate that CdS-polymer composite films have great potential for optical switching applications.

## 2. Experimental Section

### A. Sample Preparation

The CdS-polymer composite film was prepared by an ion-exchange process. The 0.13-mm-thick, free-standing Nafion™ film was first cleaned in boiling 65 – 68% nitric acid for about 30 minutes and repeatedly washed with de-ionized water until the solution was neutral at pH of 7. The Nafion™ film was then soaked in 0.5 M aqueous cadmium acetate solution for 12 hours. After repeatedly washing the Nafion film with de-ionized water to remove non-exchanged cadmium ions, the film was thoroughly dried in vacuum for 1 hour. Then the film was treated with hydrogen sulfide gas for 0.5 hours. Finally, the CdS-Nafion composite film was kept in vacuum for 3 hours in order to remove adsorbed but un-reacted hydrogen sulfide gas.

**B.     Characterization Techniques**

The morphology and size distribution of the CdS NCs in the film were inspected with a field emission transmission electron microscope (TEM, Philips FE CM300) operating at 300 kV. The crystal phase analysis was carried out on a D8 ADVANCE X-ray diffractometer (Bruker Analytical X-ray system) with glancing angle X-ray diffraction (GAXRD) configuration. The depth profile of the CdS NCs in the composite film was measured by using an ellipsometer system (J.A. Woollam Company, VASE). The transmission spectrum was measured on a UV-visible spectrophotometer (UV-1601, Shimadzu). The photoluminescence (PL) spectrum was acquired in the wavelength range of 370–900 nm at room temperature with excitation light from a He-Cd laser ($\lambda$ = 325 nm). We measured the absorption spectra before and after the pulsed laser irradiation, described below, and no measurable difference was observed, showing high photo-stability of the CdS-Nafion composite film. In addition, the transmission spectrum of a pure Nafion™ film was measured for reference. No absorbance feature, nor any interference pattern, was observed for the pure Nafion™ film in the visible wavelength range.

**C.     Nonlinear Optical Measurements**

The two-photon absorption and optical Kerr nonlinearity of the composite film were measured by using femtosecond time-resolved pump-probe and optical Kerr effect (OKE) techniques, respectively. The laser pulses used were generated by a mode-locked Ti:Sapphire laser (Tsunami, Spectra-Physics) operating at a repetition rate of 82 MHz with a pulse duration of 250 fs and a wavelength of 800 nm. A detailed description of the OKE experiment can be found elsewhere.[12] The pump irradiance in the sample was kept below 2 GW/cm$^2$, which ensured no significant influences by two-photon-created electron-hole pairs or nonlinear processes of thermal origin. The pure Nafion™ film was also examined under the same experimental conditions, and no nonlinear response was found. For reference, a 0.5-mm-thick sample of bulk CdS crystal was investigated as well under the same conditions.

## 3. Results and Discussion

### A. Characterization of CdS Nanocrystals

Figure 1 presents a TEM image of the CdS NCs in the composite film and their size distribution. By using a lognormal fit, both the average diameter of 4.1 nm and the standard deviation of 0.5 nm are inferred for the CdS NCs. By comparing to the exciton Bohr radius $a_B$ (~ 2.8 nm in bulk CdS), we shall expect strong quantum confinement effect on the optical properties of the composite film. As shown in Figure 2, the XRD pattern of the CdS-Nafion film reveals four diffraction peaks, corresponding to four signatures for the cubic (zinc blende) structural form. By fitting with Gaussian curves, the four profiles are found to be broader than their counterparts in bulk CdS. The wide band centered at 38° is attributed to pure Nafion.

Figure 3 shows a depth profile of the refractive index near one surface of the CdS-Nafion film, measured by using the ellipsometer system at 500-nm wavelength. In the composite film, there are two layers of higher refractive index with a thickness of about 0.9 $\mu$m. It is evident that the CdS NCs are formed in the areas near to the two surfaces of the free-standing composite film. And this directly results from our synthesis method. The UV-visible transmission spectrum of the CdS-Nafion film is illustrated in Figure 4a. Firstly, it displays an interference pattern in the spectral region of 500-900 nm. We attribute it to the formation of the CdS NC layers in the film. By using the Airy function for multiple beam interference from the parallel layer, we can simulate the interference pattern. Secondly, Figure 4a shows a cutoff behavior for the CdS-Nafion film, with an absorption onset located at 2.72 eV. Compared to bulk CdS (~ 2.37 eV), there is a blue shift for the CdS NCs. If the average diameter of 4.1 nm is used, we can calculate the amount of the blue shift by using the theory of Wang et al.[13], and find that it is in agreement with our experimental data from the UV-visible spectrum. In addition, the transmittance of pure Nafion™ films used in our investigation is found to be nearly 90% from 500 to 1700 nm (Figure 4b).

Figure 4a also displays the room-temperature PL emission spectrum of the CdS-Nafion film under He-Cd laser excitation at 325 nm. The PL spectrum contains two emission bands, which can be attributed to band-edge emission (440 ~ 480 nm) and defect emission (550 ~ 750 nm), respectively. The band-edge emission centered at 2.71 eV (or 458 nm) with a full width half maximum (FWHM) of 270 meV is sharper than the defect emission. For comparison, the PL spectrum of pure Nafion film was also examined and no obvious signal was observed.

**B.    Determination of Two-Photon Absorption**

To determine both two-photon absorption (TPA) and Kerr nonlinearity, we conducted the pump-probe and OKE measurements with 250-fs laser pulses of 800-nm wavelength. A decrease in the normalized transmittance of the pump-probe transient signal is shown in Figure 5a for the CdS-Nafion film. The negative change in the transmittance indicates the occurrence of TPA, which agrees with the findings by Cotter et al.[1] and Banfi et al.[3,4,6] for semiconductor-NC–doped glasses. It should be emphasized that it is different from the report by Lin et al.[11] who observed the absorption saturation in 3 wt% CdS-PMMA hybrid film with 120-fs laser pulses at 815 nm. In the absorption spectrum of the 3wt% CdS-PMMA film (Figure 1 in Ref. 11), we notice the presence of a long absorption tail up to 900 nm. It is indicative of a large amount of states in the spectral region from 500 to 900 nm for their composite film. The origins of these states could result from defects or interfacial states on the surfaces of CdS nanoparticles. Their PL spectra of the CdS-PMMA films also reveal the same fact. Hence, it is not surprising to see absorption saturation due to state-filling effects of the defect/interfacial states by photon-excited electrons or holes.

If TPA is small, the change in the transmittance, $|\Delta T|$ may be approximated to:[14]

$$|\Delta T| \propto (1-R)^3 \beta I_0 L_{eff} \exp(-\alpha L)/2\sqrt{2} \qquad (1)$$

where $I_0$ is the on-axis maximum irradiance at the focus, $\beta$ the TPA coefficient, $\alpha$ the linear absorption coefficient, $R$ the surface reflectance, and $L_{eff} = (1 - e^{-\alpha L})/\alpha$. The $\beta$ value of the CdS-Nafion film is determined by comparing with the reference sample at the zero delay time:

$$\beta_s = \beta_r \times \left(\frac{|\Delta T|_s}{|\Delta T|_r}\right) \times \left(\frac{1-R_r}{1-R_s}\right)^3 \times \left(\frac{L_r}{L_s}\right) \times \exp(\alpha_s L_s) \qquad (2)$$

The subscripts $s$ and $r$ denote the CdS-Nafion film and the bulk CdS sample, respectively. $L$ is the interaction length between the pump and probe beam over the sample. The $\beta$ value was reported to be 6.4 cm/GW for bulk CdS at 780 nm.[15] With this reference value, the magnitude of $\beta$ is determined to be 9.5 cm/GW for the CdS-Nafion film. The corresponding imaginary susceptibility ($Im\chi^{(3)}$) is $3.7 \times 10^{-12}$ esu. This TPA is at least two orders of magnitude greater than those in semiconductor-NC-doped glasses measured by Banfi *et al.*,[4] and Bindra and Kar,[7] who reported that $\beta$ is in the range of $1.2 \times 10^{-4}$ to $6 \times 10^{-2}$ cm/GW. The discrepancy is expected since our film has a larger volume fraction of CdS NCs (about 20%) whereas the volume fraction is about 0.1 ~ 0.5% in those glasses. In addition, the difference in stoichiometry is also expected to contribute to the discrepancy. The TPA cross section of $1.1 \times 10^{-47}$ cm$^4$s has been measured by Van Oijen *et al.*[16] for 5-nm-diameter CdS NCs in a polymer (PVA). For comparison, we use the definition of $\sigma = \dfrac{\beta \hbar \omega}{4 N_0}$, where $\sigma$ is the TPA cross section, $\hbar \omega$ the photon energy, and $N_0$ the density of CdS NCs in the film. By using $N_0 = 8.4 \times 10^{18}$ cm$^{-3}$, we find it to be $7.0 \times 10^{-47}$ cm$^4$s for the CdS-Nafion film, which is in the same order of magnitude as the result of Van Oijen *et al.*.[16]

## C. Determination of Optical Kerr Nonlinearity

Figure 5b displays our OKE measurement for the CdS-Nafion film. From the OKE signals at the zero delay time, the magnitude of the third-order nonlinear susceptibility $\chi^{(3)}$ of the sample is given by:[17]

$$\chi_s^{(3)} = \chi_r^{(3)} \times \left(\frac{I_s}{I_r}\right)^{1/2} \times \left(\frac{n_s}{n_r}\right)^2 \times \left(\frac{L_r}{L_s}\right) \times \left(\frac{1-R_r}{1-R_s}\right)^{3/2} \times \frac{\alpha_s L_s}{\exp\left(-\dfrac{1}{2}\alpha_s L_s\right) \times [1-\exp(-\alpha_s L_s)]} \quad (3)$$

where I is the magnitude of OKE signal and $n$ the refractive index. The subscripts *s* and *r* denote the CdS-Nafion film and the bulk CdS sample, respectively. The value $|\chi^{(3)}| = (|Re\chi^{(3)}|^2 + |Im\chi^{(3)}|^2)^{1/2}$ is calculated to be $7.4 \times 10^{-12}$ esu for bulk CdS by using the data of $n_2$ and $\beta$ reported in Ref. 15. By using the reference value, the magnitude of $\chi^{(3)}$ for the CdS-Nafion film is determined to be $3.5 \times 10^{-11}$ esu. The value of $Im\chi^{(3)}$ has been obtained from the pump-probe experiment mentioned previously. Hence, $|Re\chi^{(3)}|$ is computed to be 3.4 × 10$^{-11}$ esu, by $|Re\chi^{(3)}| = (|\chi^{(3)}|^2 - |Im\chi^{(3)}|^2)^{1/2}$, and the corresponding Kerr coefficient, $|n_2|$, is 5.7 × 10$^{-13}$ cm$^2$/W. This $n_2$ value is one order of magnitude greater than the reported value of -8.4 × 10$^{-14}$ cm$^2$/W by Lin *et al.*[11] for 3 wt% CdS NCs in PMMA. The $n_2$ value is also two orders of magnitude bigger than the measurement by Bindra and Kar[7] for semiconductor-NC-doped glasses. Again, we believe that the discrepancy is caused mainly by different volume fractions.

To evaluate the material requirements for all-optical switching devices, we calculate the one-photon and two-photon figures of merit for the CdS-Nafion film. For $\lambda$ = 800 nm, $\beta$ = 9.5 cm/GW, $n_2$ = 5.7 × 10$^{-4}$ cm$^2$/GW, $I$ = 2.0 GW/cm$^2$, and $\alpha$ = 4.7 cm$^{-1}$ measured by the ellipsometry, the figures of merit are calculated to be: $W = n_2 I/(\alpha\lambda)$ = 3.1 and $T = \beta\lambda/n_2$ = 1.3. They are close to the target values of $W > 1$ and $T < 1$.

**D.    Relaxation of Two-Photon Generated Free Carriers**

The transient response in Figure 5a is mainly dominated by the autocorrelation function of the pump and probe laser pulses, which confirms that the TPA plays a key role in the observed nonlinear absorption since TPA is an instantaneous nonlinear process. However, there is a recovery component with a characteristic time of ~ 1 ps. The exact nature of this slow component is unclear. But it is in agreement with the observation of Klimov *et al.*[18], who attributed the 1-ps component to the trapping of photo-generated holes at shallow acceptor states of CdS NCs dispersed in a glass matrix. Klimov *et al.*[18] also observed another 20~30 ps relaxation process, which they attributed to the capture of photo-excited electrons by deep centers. We did not observe it in our experiment. And it may be due to the following reasons: (1) different excitation mechanism: electron-hole pairs are generated by two-photon absorption in our case; (2) the excitation intensity is so low that the density of photo-generated electron-hole pairs is ~ $2 \times 10^{16}$ cm$^{-3}$ in our experiment; or (3) CdS NCs are embedded in polymer, which is expected to have different stoichiometry from CdS NCs dispersed in glass. The magnitude of the observed 1-ps component in our experiment is small, which confirms that the two-photo-excited free carrier (or electron-hole pairs) absorption is insignificant in our pump-probe experiment. This is consistent with the observation by Banfi *et al.*[4,6] who concluded that a negligible contribution due to photo-generated free carrier absorption is expected to the overall nonlinear absorption in CdS$_{1-x}$Se$_x$-doped glasses with pulses shorter than 200 fs.

Figure 5b reveals that the evolution of the OKE signal in the CdS-Nafion film consists of two decay processes. The fast component has the characteristic time about 200 fs, which is assigned to the instantaneous response to the laser pulse. The Kerr nonlinearity due to bound electronic effect is instantaneous. The 1-ps slow recovery process reflects the relaxation of the two-photon-excited free carriers, which could also be explained by holes' trapping.[18] It is interesting to note that the magnitude of the recovery in the nonlinear refraction is nearly three times as much as the recovery observed for the nonlinear absorption in the pump-probe experiment displayed in Figure 5a. It implies that the photo-generated free carriers' contribution to the overall nonlinear refraction is more substantial than to the overall nonlinear absorption. It is expected to be a dominant factor when semiconductor NCs are irradiated by longer laser pulses, as demonstrated by Bindra and Kar,[7] who compared the picosecond to femtosecond Z-scan measurements for the overall nonlinear refraction in CdS$_{1-x}$Se$_x$-doped glasses. It also becomes important when higher laser irradiances (~ 600 GW/cm$^2$) are used.[7]

## 4. Conclusions

In conclusion, we have measured the ultrafast and large third-order nonlinear optical susceptibilities ($Re\chi^{(3)} = 3.4 \times 10^{-11}$ and $Im\chi^{(3)} = 3.7 \times 10^{-12}$ esu) in the polymeric film embedded with highly-concentrated CdS NCs of 2-nm radius. The measurements have been conducted by using both pump-probe and OKE techniques with 800-nm, 250-fs laser pulses. The observed nonlinearities have a recovery time of ~ 1 ps with the figures of merit: W = 3.1 and T = 1.3. The two-photon-generated free carrier effects have also been observed and discussed. Our results demonstrate that CdS embedded polymeric films are promising for optical switching applications.

**Acknowledgment.** The authors thank Dr. R. Synowicki for the ellipsometrical measurements. This research is supported by the National University of Singapore (Research Grant Nos. R-144-000-104-112 and R-144-000-084-112).

**Figure Captions:**

**Figure 1.**   (a) High-resolution TEM and (b) size distribution of the CdS NCs in the composite film.

**Figure 2.**   XRD pattern of the CdS-Nafion film. The thick solid line is the measurement. The thin solid line is the fit with Gaussian curves. The de-convoluted peaks, corresponding to the individual diffraction peaks, are indicated by the dotted lines.

**Figure 3.**   Depth profile of the refractive index in the CdS-Nafion film.

**Figure 4.**   (a) UV-visible transmission spectra of CdS-Nafion film (thick solid line) and pure Nafion film (dotted line) together with the PL emission spectrum (thin solid line) of CdS-Nafion film; (b) UV-visible-IR transmission spectrum of pure Nafion film.

**Figure 5.**   (a) Pump-probe and (b) OKE signals measured as a function of the delay time for the CdS-Nafion film. The pump irradiance is 2.0 GW/cm$^2$. The symbols (filled squares) are the experimental data. The dotted lines are the autocorrelation function of the laser pulses. The solid lines are the best fits based on two exponentially decay terms.

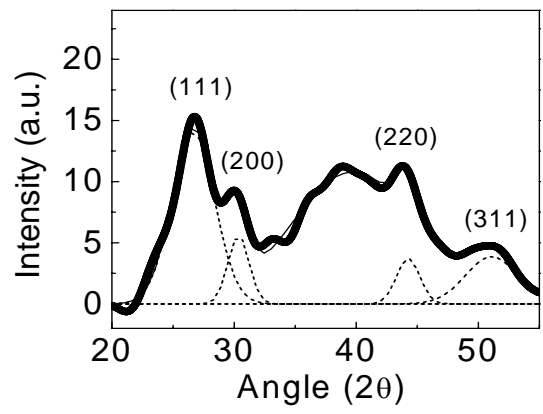

**Fig. 2 J. He et al.**

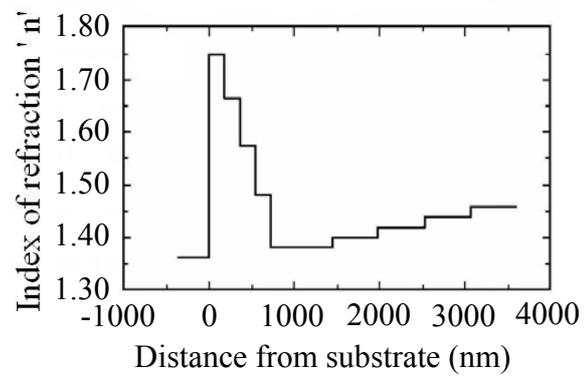

**Fig. 3 J. He et al.**

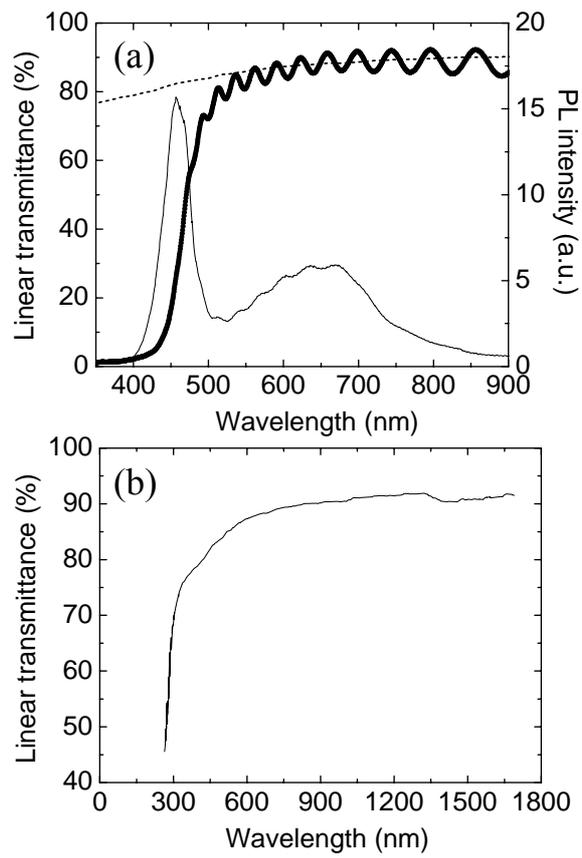

**Fig. 4 J. He et al.**

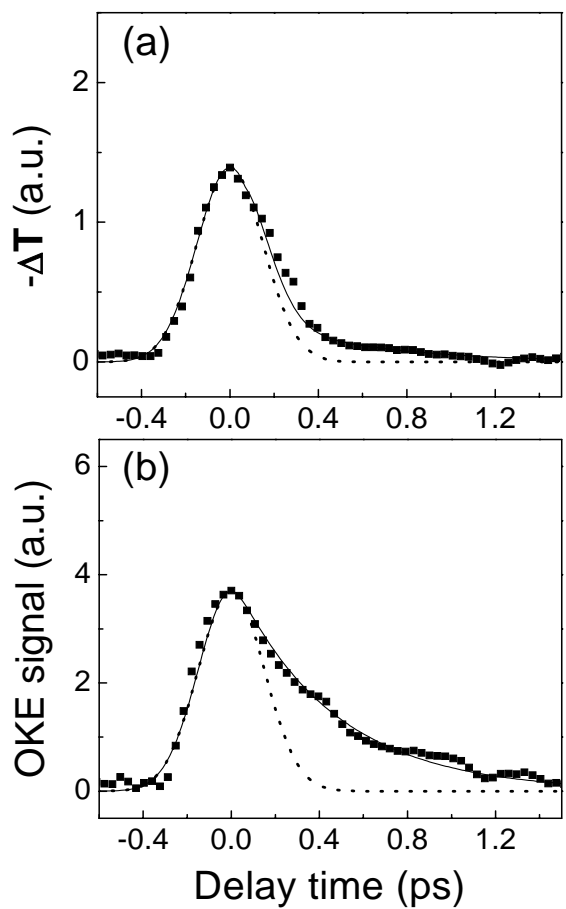

**Fig. 5 J. He et al.**